\documentclass[preprint2]{aastex631}

\begin{document}

\title{SuNeRF: 3D reconstruction of the solar EUV corona using Neural Radiance Fields}

\author[0000-0002-9309-2981]{Robert Jarolim}
\affiliation{University of Graz, Universit\"atsplatz 5, 8010 Graz, Austria}

\author[0000-0002-5181-7913]{Benoit Tremblay}
\affiliation{High Altitude Observatory,
    3080 Center Green Dr.,
	Boulder, CO 80301, USA}

\author[0000-0002-4716-0840]{Andr\'es Mu\~noz-Jaramillo}
\affiliation{Southwest Research Institute,
	1050 Walnut St., Suite 300,
	Boulder, CO 80302, USA}

\author[0000-0003-3875-7757]{Kyriaki-Margarita Bintsi}
\affiliation{Imperial College London, London SW7 2AZ, UK}

\author[0000-0002-9888-6262]{Anna Jungbluth}
\affiliation{European Space Agency (ESA) - ECSAT, Fermi Avenue, Harwell, UK}

\author{Miraflor Santos}
\affiliation{Massachusetts Institute of Technology,
	77 Massachusetts Ave,
	Cambridge, MA 02139, USA}

\author[0000-0002-8164-5948]{Angelos Vourlidas}
\affiliation{Johns Hopkins University Applied Physics Laboratory,
	11100 Johns Hopkins Rd.,
	Laurel, MD 20723, USA}

\author[0000-0002-3783-5509]{James P. Mason}
\affiliation{Johns Hopkins University Applied Physics Laboratory,
	11100 Johns Hopkins Rd.,
	Laurel, MD 20723, USA}

\author[0000-0002-6648-0591]{Sairam Sundaresan}
\affiliation{Intel Labs,
	2200 Mission College Blvd.,
	Santa Clara, CA 95054, USA}

\author[0000-0003-1759-4354]{Cooper Downs}
\affiliation{Predictive Science Inc.,
        9990 Mesa Rim Rd. Suite 170,
        San Diego, CA 92121}

\author[0000-0002-2633-4290]{Ronald M. Caplan}
\affiliation{Predictive Science Inc.,
        9990 Mesa Rim Rd. Suite 170,
        San Diego, CA 92121}

%% Note that the \and command from previous versions of AASTeX is now
%% depreciated in this version as it is no longer necessary. AASTeX 
%% automatically takes care of all commas and "and"s between authors names.

%% AASTeX 6.31 has the new \collaboration and \nocollaboration commands to
%% provide the collaboration status of a group of authors. These commands 
%% can be used either before or after the list of corresponding authors. The
%% argument for \collaboration is the collaboration identifier. Authors are
%% encouraged to surround collaboration identifiers with ()s. The 
%% \nocollaboration command takes no argument and exists to indicate that
%% the nearby authors are not part of surrounding collaborations.

%% Mark off the abstract in the ``abstract'' environment. 
\begin{abstract}

To understand its evolution and the effects of its eruptive events, the Sun is permanently monitored by multiple satellite missions. The optically-thin emission of the solar plasma and the limited number of viewpoints make it challenging to reconstruct the geometry and structure of the solar atmosphere; however, this information is the missing link to understand the Sun as it is: a three-dimensional evolving star.
We present a method that enables a complete 3D representation of the uppermost solar layer (corona) observed in extreme ultraviolet (EUV) light. We use a deep learning approach for 3D scene representation that accounts for radiative transfer, to map the entire solar atmosphere from three simultaneous observations. We demonstrate that our approach provides unprecedented reconstructions of the solar poles, and directly enables height estimates of coronal structures, solar filaments, coronal hole profiles, and coronal mass ejections.
We validate the approach using model-generated synthetic EUV images, finding that our method accurately captures the 3D geometry of the Sun even from a limited number of 32 ecliptic viewpoints ($|\text{latitude}| \leq 7^\circ$). We quantify uncertainties of our model using an ensemble approach that allows us to estimate the model performance in absence of a ground-truth.
Our method enables a novel view of our closest star, and is a breakthrough technology for the efficient use of multi-instrument datasets, which paves the way for future cluster missions.

\end{abstract}

%% Keywords should appear after the \end{abstract} command. 
%% The AAS Journals now uses Unified Astronomy Thesaurus concepts:
%% https://astrothesaurus.org
%% You will be asked to selected these concepts during the submission process
%% but this old "keyword" functionality is maintained in case authors want
%% to include these concepts in their preprints.
% \keywords{Classical Novae (251) --- Ultraviolet astronomy(1736) --- History of astronomy(1868) --- Interdisciplinary astronomy(804)}

%% From the front matter, we move on to the body of the paper.
%% Sections are demarcated by \section and \subsection, respectively.
%% Observe the use of the LaTeX \label
%% command after the \subsection to give a symbolic KEY to the
%% subsection for cross-referencing in a \ref command.
%% You can use LaTeX's \ref and \label commands to keep track of
%% cross-references to sections, equations, tables, and figures.
%% That way, if you change the order of any elements, LaTeX will
%% automatically renumber them.
%%
%% We recommend that authors also use the natbib \citep
%% and \citet commands to identify citations.  The citations are
%% tied to the reference list via symbolic KEYs. The KEY corresponds
%% to the KEY in the \bibitem in the reference list below. 

\section{Introduction}

% observation of the Sun
The outermost layer of the solar atmosphere, the corona, is highly structured by its magnetic field. Observations in Extreme Ultraviolet (EUV) light allow us to study this structure and its evolution in the low corona in an unprecedented way. Several space missions image the solar atmosphere in specific wavelength bands in an effort to understand the Sun and its effects on Earth. The Solar Terrestrial Relations Observatory \citep[STEREO: ][]{kaiser2008stereo} consists of two satellites on slightly displaced 1-AU orbits, which cause a steady drift of $22.5^\circ$/year of the two satellites relative to Earth. By adding the Solar Dynamics Observatory \citep[SDO: ][]{pesnell2012sdo}, which observes the Sun from Earth's perspective, the Sun can be simultaneously observed from three viewpoints.

% short-time scale evolution
On short-time scales, the magnetic field evolves into highly-structured localized regions \citep[active regions: ][]{driel2015active_regions} that give rise to solar eruptions and ejections of plasma into interplanetary space. Both are a direct risk for our space assets \citep{temmer2021space_weather}. The scientific understanding of pre-eruptive structures is key for advancing our ability to predict eruptions, and yet it remains elusive \citep[e.g.][]{Patsourakos_etal2020}. This was one of the motivations and early science targets of the STEREO mission \citep[e.g.][]{Aschwanden_etal2009}. With two vantage points, the 3D triangulation of specific spatial points is possible from imaging data \citep{inhester2006stereoscopy, aschwanden2008first, Liewer2009SoPh..256...57L, bemporad2009ApJ...701..298B}. However, for the optically thin medium of the solar atmosphere, assigning observed intensities to a specific spatial point leads to ambiguous results due to the effects of integration along the line-of-sight \citep[e.g.,][]{Aschwanden_2011}. In other words, the intensity at each pixel is the integral of all emitting and absorbing plasma along the line-of-sight. Consequently, the idealized view of a single emitting or solid point is insufficient to reconstruct the solar atmosphere. Tomographic reconstructions provide a different approach by reconstructing the 3D temperature and density profile of the solar EUV corona, utilizing multi-viewpoint observations and the solar rotation \citep{franzin2009demt, vasquez2009multi_tomography, Kramar2014, Cho2020}. This approach has typically a temporal resolution of about 14 days and is not suited to study regions characterized by fast dynamics \citep{vasquez2016corona_3d}.

% long-time scale evolution
On long-time scales, the Sun evolves over an 11-year cycle, which is distinctly marked by global changes \citep{hathaway2015solar_cycle}. During solar minimum, the magnetic field is mostly poloidal, with a low number of active regions clustered near the solar equator. In this period, coronal holes, which enclose open field line configurations, are present at the solar poles \citep{cranmer2009coronal_holes} and can extend down to low-latitudes. With the transition to the maximum of the cycle, and the corresponding reconfiguration to a toroidal magnetic field, active regions emerge at high latitudes and the polar coronal holes vanish. During the maximum phase, mostly small coronal holes appear at low-latitudes, and the number and complexity of active regions drastically increases. These phenomena are a direct effect of the solar dynamo and evolving magnetic field.

% a complete image of the Sun
To relate both small and large-scale changes to the physical process that generate the magnetic field of the Sun, and consequently of other Sun-like stars, a complete picture of the Sun and its atmosphere is required. A frequently applied approach are synoptic maps which use observations from only one instrument and make use of the solar rotation period of 27 days. Here, an $\sim 1^\circ$ central slice is consecutively extracted to complete the image over the course of one full solar rotation. The observed slices are then reprojected onto the surface of a sphere to obtain a 2D image (Fig. \ref{fig:synchronic}). A similar approach results in synchronic maps, which combine simultaneous (or nearly simultaneous) observations from different satellites located at different positions into a single image of the Sun \citep[e.g.][]{caplan2016}.

% shortcomings of current state-of-the-art
Both approaches have significant shortcomings that primarily affect extended and/or long-lived structures due to their temporal evolution, and line-of-sight overlapping \citep[e.g., lifetime of days-weeks for small active regions; ][]{driel2015active_regions}. Recent space-based observations originate in the ecliptic plane (latitudes of approx. $\pm 7^{\circ}$ or less), and are strongly limited in terms of the number of simultaneous observations. These constraints have consequences for the spherical assumption, where extended features (e.g., coronal loops) become warped towards the solar limb. While there are observing periods where a full coverage of the entire Sun exists (2011-2014), the observations of the poles are limited due to the limiting viewing angle of the ecliptic-orbiting telescopes. Only the combination of simultaneous imaging from three (or more) viewpoints can provide the maximum information possible for this difficult to access location. 
% Other points to mention: 
% (...) assumption that the observed intensity originates from a single plane and neglecting the three-dimensionality of the solar corona. This leads to projection effects that are especially prominent near the limb (i.e., edge) of the solar disk and reduces the scientific utility of off-limb imaging.
% Synchronic maps: Inter-calibration of instruments.

% method
To address these challenges, we devised a novel approach to provide a complete 3D reconstruction of the global solar EUV corona.
Our method builds upon a state-of-the-art deep learning method for 3D scene representations \cite[Neural Radiance Fields (NeRFs): ][]{mildenhall2021nerf}. The goal is to reconstruct the 3D geometry of a scene from a set of images and their known viewpoints. For a sufficiently large dataset (e.g., about dozens of images), the underlying 3D representation can be recovered and images from any viewpoint can be rendered. We adapt this method to match the physical reality of the Sun, using a simplified radiative transfer approach and a geometric sampling approach to create a solar radiance field Sun Neural Radiance Fields (SuNeRFs; see Sect. \ref{sec:method}) that, accounts for the line-of-sight ambiguity of an optically-thin source \citep[c.f.,][]{bintsi2022sunerf}. In addition, we account for insufficient coverage and dynamic changes by including a time component into our model.

% transition to results
SuNeRFs can be deployed to consistently combine observations from multiple vantage points into a 3D representation and then render observations at arbitrary viewpoints. In addition, this approach goes one step further and directly enables height estimates of coronal structures, solar filaments, coronal hole profiles, and solar eruptive events.

\section{Method} 
\label{sec:method}

% Introduction
To reconstruct the 3D solar atmosphere, we consider both simulated data and observation sequences from SDO/AIA and STEREO/EUVI (A and B) in specific wavelength channels. 
Our goal is to obtain a scene representation from these image sets and known viewpoints \citep[c.f.,][]{mildenhall2021nerf}. From the resulting representation we can render novel viewpoints, extract height information, and analyze the plasma distribution in terms of emitting and absorbing material.

\subsection{Simulated data}
\label{sec:data_simulated}

To validate our method, we leverage a magnetohydrodynamic (MHD) simulation, developed by Predictive Science Inc. (PSI), of the solar corona used to forecast the state of the solar atmosphere prior to the 2019-07-02 total solar eclipse \citep[see the website\footnote{Predictive Science Inc. 2019-07-02 total solar eclipse prediction: \href{https://www.predsci.com/eclipse2019}{www.predsci.com/eclipse2019}.} and][for more details]{Boe2021}. The data set consists of a single snapshot from which we extract full-disc images from 256 viewpoints at a resolution of $1024 \times 1024$ pixels (2.63 arcsec per pixel) in three different wavelengths (171 $\text{\AA}$, 193 $\text{\AA}$, 211 $\text{\AA}$). We pre-process the simulated data analogously to the observational data introduced in Sect. \ref{sec:data_obs}.

\subsection{Observational data}
\label{sec:data_obs}

From 2010 to 2014, three satellites capable of imaging in similar EUV channels were orbiting the Sun, namely the Solar Dynamic Observatory (SDO) equipped with the Atmospheric Imaging Assembly instrument \citep[AIA: ][]{lemen2012aia}, and the twin Solar Terrestrial Relations Observatory (STEREO) spacecraft each equipped with a EUV Imager \citep[EUVI: ][]{wulser2004euvi}. 
All three satellites orbit in the ecliptic plane, with SDO in geosynchronous orbit and the STEREO satellites progressively advancing  (STEREO-A) and trailing  (STEREO-B)  Earth. When the separation angle between SDO and each of the STEREO spacecraft exceeded $90^\circ$, it allowed to view the full Sun at one instant, for the first time. 

% ITI
An important issue in the context of this project is the design, performance, and operational differences among the various instruments. We apply Instrument-To-Instrument translation \citep[ITI:][]{jarolim2022instrument} to translate full-disk observations from STEREO to the SDO domain. With ITI, we can account for differences in resolution, calibration, and filter bands between the individual instruments. 
The AIA and EUVI instrument have common observing wavelengths (171 and 304 {\AA}). For the wavelengths 195 and 284 {\AA}, we use ITI to approximate the 193 and 211 {\AA} channels of AIA \citep[see][]{jarolim2022instrument}. Although labeled as single wavelength channels, the SDO/AIA and STEREO/EUVI spectral channels actually result from spectral integration as specified by the spectral response of the instruments and wavelength filters \citep{doi:10.1126/sciadv.aaw6548}.
The multi-channel translation of ITI is a pre-requirement to have an appropriate approximation to combine homogeneous observations of this channel. The resulting homogenized dataset serves as approximation of multi-viewpoint SDO/AIA observations in four spectral bands (171$\text{\AA}$, 193$\text{\AA}$, 211$\text{\AA}$, 304$\text{\AA}$).

We resample all data to 1.2 arcsec per pixel. The focal length $f$, as defined by our ray tracing method (discussed below), is then accordingly computed individually for each observation:
\begin{equation}
    f = \frac{W / 2}{\arctan(1.2 \cdot W / 2)},
\end{equation}
where $W$ refers to the width of the image. Note that this focal length is computed from the satellite position and observed image, and is not directly related to the actual focal length of the instrument, due to the adjustment of the image size and resolution. Furthermore, we only consider square images. In this context, we analyze the propagation of diverging rays from the observer, with the parameter $f$ specifying the angle of view. This approach departs from the assumption of small angles, where rays would be considered parallel.
%With the time of the observation, distance from Sun, Carrington latitude and longitude, we fully specify the observer.
For each observer, we specify the time of the observation, the distance from the Sun, and the Carrington longitude and latitude. 

% The Solar Orbiter \cite[SolO; ][]{mueller2020solo} mission is equipped with an EUV instrument that also provides observations of the 304 {\AA} channel. We use the observation for a qualitative comparison, without further inter-calibration with the AIA reference. We note that an additional calibration is required to utilize this data for training.

\begin{figure}
    \centering
    \includegraphics[width=\linewidth]{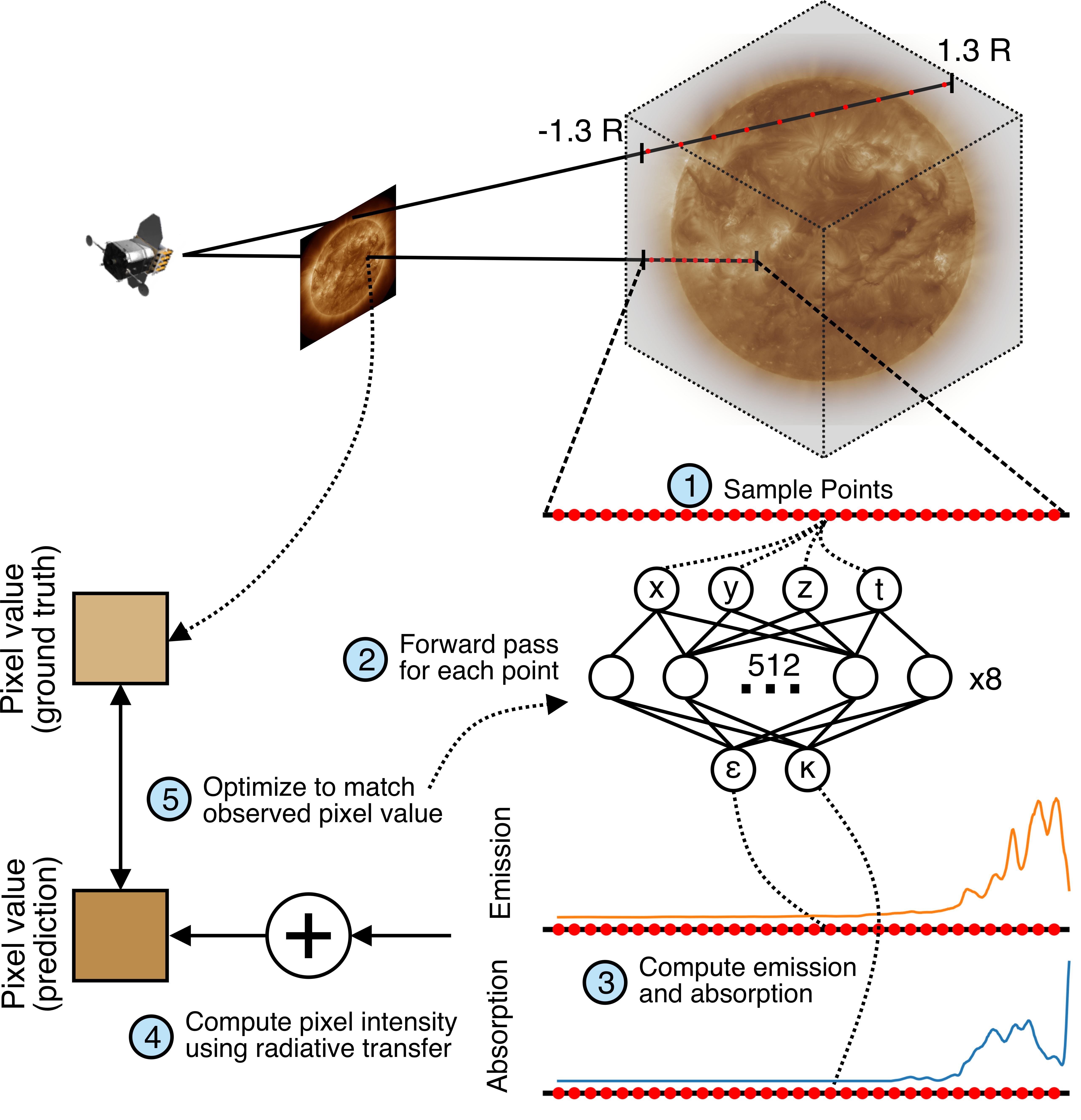}
    \caption{Overview of the proposed method for 3D reconstruction of the solar EUV corona. Our method consists of five steps. (1) For each pixel in the input video sequence we sample points along the ray path, ranging from -1.3 to 1.3 solar radii. The endpoints of rays passing through the Sun are fixed to the solar surface, to account for its opaqueness. (2) The sampled points are passed through the neural network, which outputs the emission $\epsilon$ and absorption $\kappa$ coefficient per point. (3) For each ray we construct the emission and absorption along the ray. (4) The total observed intensity is then computed by integrating all sampled points, where each intensity value is reduced by absorption along the path of propagation (from the origin to the observer). (5) The predicted intensity value is then compared to the actual pixel value, which serves as loss function for our model training. Per update step we optimize a set of 32,768 rays. By iteratively fitting all pixel values, we obtain a complete spatial and temporal representation of the solar corona.}
    \label{fig:method}
\end{figure}

% Method introduction
% In Fig. \ref{fig:method}, we provide an overview of our approach. We use a neural network that acts as mesh-free representation of the modeled volume. Therefore, our goal is to obtain a single reconstruction for a given video sequence of observations. For each image pixel we sample points along the line-of-sight. The points $(x, y, z, t)$ serve as input to our fully-connected SInusoidal REpresentation Network \citep[SIREN: ][]{sitzmann2020siren}, that maps each point over eight hidden layers with 512 neurons each, to the corresponding emitted and absorbed intensity ($\epsilon$, $\kappa$). Using ray tracing methods, we combine the emission and absorption along the ray to an intensity value. Since the rendering process is fully-differentiable, we can use the observed intensity value as reference to fit our model, using standard backpropagation.

\subsection{Model Training}

% approach
We adapt the approach from \citet{mildenhall2021nerf}. We use a neural network to represent the 3D scene and ray tracing methods for learning the representation. Information of the entire scene is stored in the weights of the neural network, rather than having an explicit grid representation. To be precise, we train a neural network for each individual scene, in contrast to classical deep learning applications which learn to perform a general task. 

% ray tracing
For each image pixel, we determine the light ray path using known information from the observer position and viewing angle. Along this path, we sample points ($x, y, z$) within the simulation volume and use classical ray tracing methods to compute the integrated pixel value. In the NeRF approach proposed by \citet{mildenhall2021nerf}, each point is associated with a color and density. The pixel value is computed by the weighted sum of colors of the sampled points along the ray, where the weights are the cumulative sum of the density values. In other words, this approach computes the likelihood of a ray being absorbed along the line-of-sight. The training images, together with the observer position, serve as target for model training. By iteratively updating the neural network to match the pixel values of the training images, we obtain the underlying scene representation. This approach has shown the ability to obtain photorealistic images at interpolated positions \citep{mildenhall2021nerf, li2022video_nerf}.

% SuNeRF
In contrast to a classical NeRF, where the volume is composed of colors and densities, the solar atmosphere consists of emitting and absorbing plasma. In addition, NeRFs require dozens of images in order to learn the scene representation, while only three simultaneous observations are available for observations of the solar EUV corona. Therefore, we apply four primary changes to the NeRF approach. (1) We replace density and color with wavelength-channel-specific emission and absorption. These capture how plasma with temperatures observed by a given wavelength channel emit and absorb light. Each pixel value refers to the total intensity that is observed along the line-of-sight for a given wavelength channel and is expressed in units of Data Number per second (DN/s). (2) We utilize the temporal evolution and solar rotation to obtain more viewpoints \citep[c.f., ][]{frazin2005rotational_tomography}. From this approach, we obtain a full scan of the Sun per observing instrument within one solar rotation ($\sim$ 27 days). For the definition of viewpoints, we consider all observations in heliographic Carrington coordinates, which are solar centric. We explicitly include time in the query coordinates of the NeRF to account for a consistent treatment of dynamic changes in the solar atmosphere. (3) We account for the known geometry of the Sun by only sampling points up to the solar surface. With this, we avoid an invalid placement of emitting plasma from below the solar surface. The change to monochromatic intensity values, instead of colors, further implies that there is no need for a background \cite[c.f.,][]{mildenhall2021nerf}. (4) An optional component for NeRFs is a viewing-angle dependence of the pixel color. For our implementation, we do not consider any view dependence of the observed intensities.

% model architecture
In Fig. \ref{fig:method}, we provide an overview of our approach. For our 3D representation, we use a fully-connected SInusoidal REpresentation Network \citep[SIREN: ][]{sitzmann2020siren}, with eight layers where each contains 512 neurons, resulting in about 2M free parameters. Our model takes four input coordinates ($x, y, z, t$) and outputs the estimated emission $\epsilon(x, y, z, t)$ and absorption $\kappa(x, y, z, t)$ coefficient for a given wavelength. We use an exponential and ReLU activation (setting negative values to zero) for the emission and absorption coefficients, respectively. With this we obtain positive values for both coefficients. We then compute the total emission $E$ at each point by multiplying with the spacing of the sampled points $ds$,
\begin{equation}
    E = \epsilon \cdot ds.
\end{equation}
Each emitted intensity value propagates through the line-of-sight to the observer, where absorption can reduce the intensity value at points between the observer and the point of emission. We model absorption values between [0, 1], where 1 refers to full transmission and 0 to total absorption. We compute the absorption $A$ for each sampled point based on the spacing,
\begin{equation}
    A = \exp{(- \kappa \cdot ds)}.
\end{equation}
The total intensity $I$ is then computed as the sum of the absorption reduced emissions,
\begin{equation}
    I = \sum_k^N E_k  \prod_{i}^{k-1} A_i,
\end{equation}
where $i$ and $k$ refer to the points along the sampled ray, and $N$ to the total number of sampled points along the ray. Therefore, each emission $E$ is reduced by all absorption values $A$ in the range from the emitting point to the observer. We note that the first point is always transmitted without absorption ($A_{1} = 1$), and the last sampled absorption point has no effect.

% adjustment of encoding
For all computations,  the spatial scales are normalized to 1 $R_{\odot}$. We use positional encoding, which implies periodic boundary conditions at $2\pi$. Therefore, we scale coordinates to units of 2 $R_{\odot}$ prior to the encoding (simulation volume of $[-2\pi, 2\pi]$  $R_{\odot}$) to avoid reflections at the boundary.

% data normalization
For training, we adjust the pixel-value range based on the average maximum value \citep[c.f.,][]{jarolim2022instrument} and use an asinh stretch to mitigate the contributions by bright pixels. To account for this scaling, we use the same stretch function when computing the final pixel values ($I_{\rm scaled}$) of our rendered images,
\begin{equation}
    \label{eq:stretch}
    I_{\rm scaled} = \frac{\text{asinh}(\widehat{I} / a)}{\text{asinh}(1 / a)},
\end{equation}
with $a = 0.05$.

% sampling
We apply two sampling strategies to prioritize points from the most important regions. (1) We only sample from -1.3 to +1.3 $R_{\odot}$ with respect to the solar center, such that the default sampling range spans 2.6 $R_{\odot}$. In addition, we stop the sampling where light rays intersect the solar surface (1 $R_{\odot}$), such that a ray that is pointing at the center of the solar disk is sampling a range of 0.3 $R_{\odot}$. Therefore, the total line-of-sight is dependent on the viewing direction. We note that this sampling is independent of the observer distance. (2) Similarly to \citet{mildenhall2021nerf}, we use a coarse and a fine network. For the coarse network, we sample 64 points uniformly distributed along the ray. The emission values of the coarse network are then used as a probability distribution to sample 128 additional non-uniformly-distributed sampling points. These points and the 64 uniformly-sampled points are then used as input in the fine network which serves as the primary training objective. For all our evaluations we use the outputs of the fine network.

For our model training, we select random rays from the full set of image pixels, where the number of rays per batch is adjusted to the available computational resources. For our reconstructions, we consider each wavelength separately. Therefore, our model is trained with observations from a single wavelength channel.

We initiate the model training with a centered crop of the image, where we use pixels within [-1000, 1000] arcsec in Helioprojective coordinates. We train on this data set for 1 epoch ($\sim$100,000 iterations) to prevent divergence at the beginning of the model training and to emphasize the reconstruction of the solar disk in the initial phase of training. Model training is then continued with the full field of view of the instruments.

When training for smaller field-of-views (subframes), we use a model trained on the full field of view as an initial starting point. This limits divergences and already incorporates the global geometry of the Sun. In the second training step, we crop subframes at a fixed coordinate point from the full-disk observations, which serves as our new training set. Therefore, pixels outside of the region of interest are neglected and are no longer suitable for further analysis after fitting the model to the subregion. The reduced spatial information that the model needs to fit enables the increase of temporal and spatial resolution (Sect. \ref{sec:eruption}).

% monitor training
During model training, we separate one image of the training set to validate performance. We monitor the peak-signal-to-noise-ratio (PSNR), mean-squared-error (MSE) and structural-similarity-index \citep[SSIM: ][]{wang2004ssim}. For all our model runs, we found a monotonic performance increase, and a similar performance independent of the initialization. We note that starting with a field-of-view that is too large can result in irreversible divergence within the first epoch.

% fine coarse
During each training step, we optimize both the coarse and fine model to match the target pixels. We use MSE as loss function.

% Absorption regularization
We introduce an additional regularization of the modeled absorption that reduces nonphysical results above the poles. Viewpoints from the ecliptic provide no observations of the solar surface and therefore only model low intensity values. This allows for arbitrary absorption values at these regions. We add an additional loss factor $\mathcal{L}_{\rm regularization}$ to our optimization that suppresses absorption values $\kappa$ above 1.2 $R_{\odot}$ to stronger penalize large absorption at greater heights:
\begin{equation}
    \mathcal{L}_{\rm regularization} = \max \{d - 1.2, 0\} \times (1 - \kappa),
\end{equation}
where $d$ refers to the distance from the solar center in $R_{\odot}$. When training with full field of view images, we add the regularization to MSE loss, whereas we set the regularization to 0 when training with subframes (i.e., no polar regions).

\subsection{Model Evaluation}
\label{sec:model_evaluation}
% Quality Estimation
To evaluate our method, we compute the PSNR and SSIM between our reconstruction and the ground-truth image. Here, we compare scaled images (c.f., Eq. \ref{eq:stretch}), to avoid that bright pixels dominate our evaluation.

% Uncertainty Estimation
To provide an uncertainty estimation, we train five independent models with randomly intialized weights. To generate uncertainty maps, we render the same viewpoint per model and compute the standard deviation of the resulting set of images. Therefore, for regions that are less constrained by observations (e.g., single viewpoint), different solutions can be found. This degree of freedom is then reflected by our uncertainty metric.

% Height Estimation
Calculating height from imaging data is challenging since there exists no exact spatial point that can be associated with an observed intensity value. We determine height information by computing the average distance from the solar center for all sample points along the ray, using channel-specific emission as a weight. In other words, for each ray we compute the height of the average emission value. This implies that dark regions (e.g., coronal holes, filament channels) appear as elevated structures, since the primary emission is coming from higher layers, and should not be considered for mapping the spatial location of the solar feature.

% Emission Profiles
For the visualization of 2D slices of the solar atmosphere, we sample radial points along a given longitude. We pass each point through our network to obtain the respective channel-specific emission and absorption coefficients. For the emission we use an asinh stretch, analogous to Eq. \ref{eq:stretch}.

% baseline
As a baseline approach for rendering novel viewpoints, we use a simple reprojection. More specifically, we combine observations at a given time from STEREO-A, -B and SDO into a synchronic (Carrington) map. In case of overlapping regions, we compute the mean value of the intersecting pixels. For missing regions (i.e., polar regions that are not covered by any instrument), we use the mean value over the entire image as a filling value. To synthesize novel viewpoints, we project the map onto a unit sphere, and rotate to the perspective to the new viewpoint.

This approach has several shortcomings. By projecting to a 2D map, we lose the off-disk information, and consequently this method can not account for the faint emission that is observed over the solar limb. This further approximates the atmosphere as flat, neglecting the height information when projecting to new viewpoints. This is especially problematic for extended structures (e.g., coronal loops), where projection effects become especially prominent closer towards the limb. Finally, combining overlapping regions is problematic. Due to the projection effects, the observations of the individual instruments do not align pixel-wise in the synchronic map. Therefore, computing the mean of the overlapping pixels, leads to blurred features.

Here, we only consider a basic approach as our baseline, but note that there are more advanced data homogenization methods \citep[e.g.,][]{hamada2020homogenization} and image stitching methods \citep[e.g.,][]{caplan2016} available.

\section{Results}
\label{sec:results}

We validate our method against simulations (Sect. \ref{sec:simulation}) and observations (Sect. \ref{sec:poles} and \ref{sec:height}).
Additionally, we rely on our method's ability to capture the volume's temporal evolution to reconstruct a solar eruption at high temporal cadence (Sect. \ref{sec:eruption}).
%Our method is able to capture the temporal evolution within the 3D volume. We use this to reconstruct a solar eruption with high temporal resolution (Sect. \ref{sec:eruption}). 

\subsection{Validation with simulated EUV observations}
\label{sec:simulation}

To validate our method we use the simulated data introduced in Sect. \ref{sec:data_simulated}. For our model training, we consider each wavelength separately, and use observations viewed from $|\text{latitude}| \leq 7^\circ$ from the ecliptic (i.e., 32 images). This constraint matches the observing capabilities of the SDO and STEREO satellites, that are solely located on the ecliptic plane. Our data set corresponds to a single snapshot, therefore we consider a static atmosphere where we set the temporal coordinate to zero. For our evaluation we focus on the comparison of ground-truth and SuNeRF reconstructed images. With this we assess the global plasma distribution, and can estimate the model performance with standard image metrics (i.e., PSNR, SSIM).

In Table \ref{table:evaluation} we summarize the quantitative evaluation of a baseline spherical reprojection  and SuNeRF reconstructions. We compute the peak-signal-to-noise-ratio (PSNR), structural-similarity index (SSIM), relative mean-absolute-error (MAE) and relative mean-error (ME), based on images from non-ecliptic viewpoints that are not seen by the model during training (i.e. a test set of 224 images) and the synthesized SuNeRF images. As baseline we use a standard reprojection of synchronic maps, where each point is mapped to a spherical surface (Sect. \ref{sec:model_evaluation}). We note that this method can not account for the off-limb emission of the Sun, while our SuNeRF method can take the extent of the 3D corona into account. The evaluation shows that our method provides a clear improvement over the baseline approach. Moreover, the metrics averaged over all considered wavelength channels indicate a strong similarity to the ground truth, with a MAE of $0.5\%$, SSIM of 0.97 ,and PSNR of 40.4 dB, which throughout outperforms the baseline with a MAE of $3.2\%$, SSIM of 0.73 ,and PSNR of 22.6 dB. The ME indicates that there is no systematic over-, or under-estimation (max ME: $-0.3\%$). 

\begin{figure*}
    \centering
    \includegraphics[width=\linewidth]{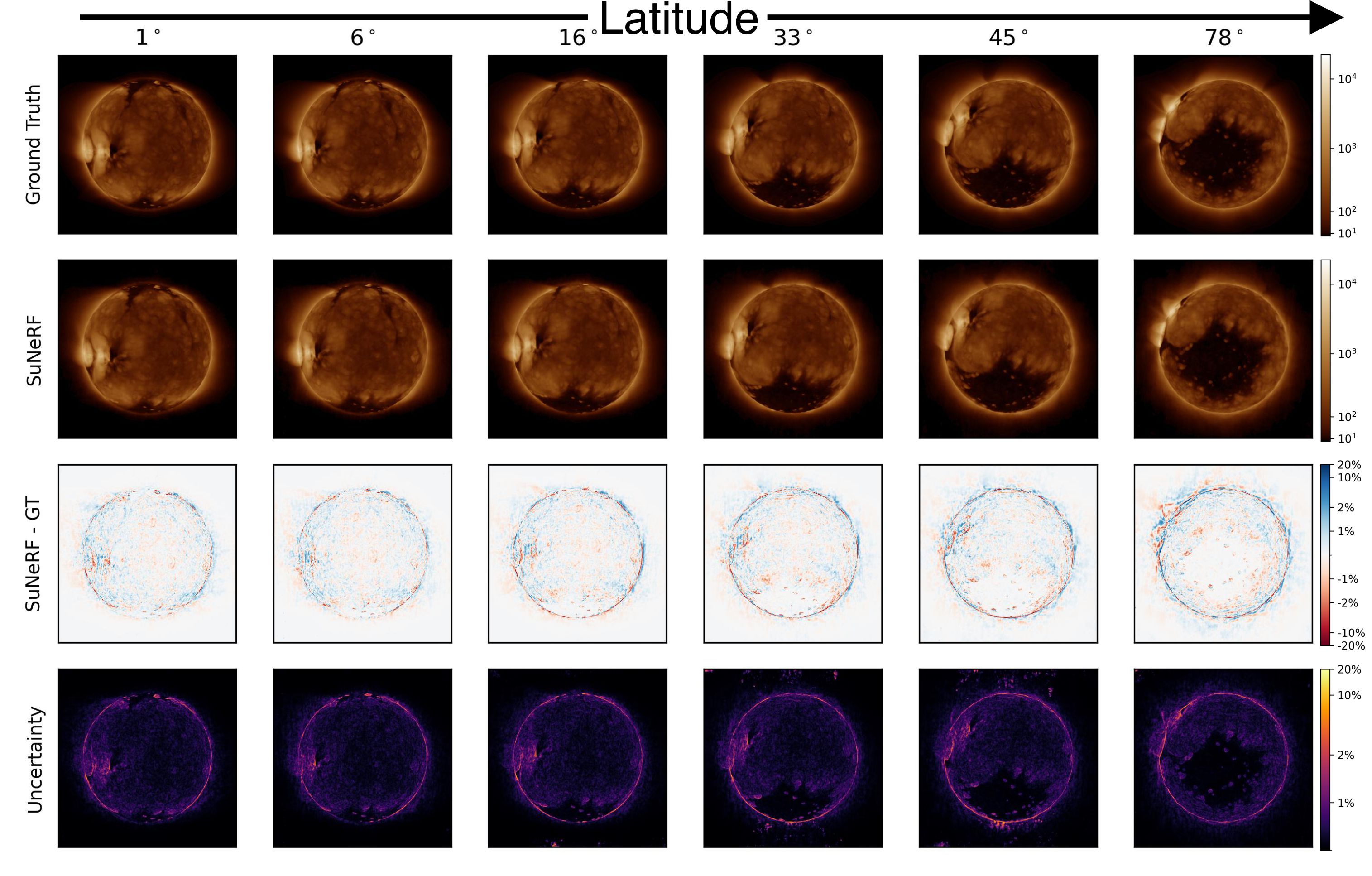}
    \caption{Qualitative evaluation of the SuNeRF reconstruction of simulation data. We compare the ground truth image (first row) with our SuNeRF reconstructions (second row) at different latitudes. The pixel-wise difference shows where our reconstruction deviates from the ground-truth reference (third row). The uncertainty maps are computed from an ensemble of five different runs (fourth row). The SuNeRF reconstructions appear indistinguishable at larger scales. The primary errors occur close to the off-limb and due to pixel-wise shifts, while large-scale structures are consistent. The uncertainty maps align with regions of increased error, which enables a performance estimate in the absence of a reference image.}
    \label{fig:simulation}
\end{figure*}

In Fig. \ref{fig:simulation} we compare ground truth and reconstructed EUV images at different latitudes. Our method results in almost identical images to the ground truth. The difference maps indicate that errors primarily originate from off-limb features, and pixel-wise shifts. In addition, there is only a small quality decrease with increasing latitudes (e.g., PSNR from 46 to 41 dB for 193 $\text{\AA}$), which suggests a valid approximation of the 3D geometry of the solar corona. This is also visible from the pixel-wise error maps. 
Our SuNeRF model achieves the highest similarity in quiet Sun regions as well as the interior of coronal holes. On the other hand, features such as active regions are more complex in addition to spanning a larger range of emission values. This results in larger differences when compared to the ground truth. Larger errors are expected close to the limb as the model integrates over a larger range of uncertainties. Therefore, independent of the viewing angle, inferences on the limb will have an increased uncertainty.

To further estimate model errors, we render images from an ensemble \citep{lakshminarayanan2017simple} of five SuNeRFs with different random weight initializations and compute the standard deviation per pixel (Fig. \ref{fig:simulation}). The resulting uncertainty maps show a good spatial overlap with the pixel-wise difference  maps. We perform a pixel-wise comparison of the uncertainty estimates and the mean-absolute-error over the full test set (see also App. \ref{sec:extended} for pixel-wise scatterplots), where we find a Pearson correlation of 0.63 and a Spearman correlation of 0.69.  This indicates that the ensemble approach reflects the model errors and can serve as an error estimate in the absence of a ground truth.

\begin{table}[t]
\caption{Summary of the quantitative evaluation of SuNeRFs and baseline (i.e. reprojected synchronic maps) methods. We evaluate both methods on the test dataset (224 non-ecliptic viewpoints), where we compare the peak signal-to-noise-ratio (PSNR), structural similarity index (SSIM), relative mean absolute error (MAE), and relative mean error (ME). }             % title of Table
\label{table:evaluation}      % is used to refer this table in the text
\centering                          % used for centering table
\begin{tabular}{| l || c c c c |}        % centered columns (4 columns)
\hline
Method & PSNR & SSIM & MAE & ME  \\    % table heading 
\hline                        % inserts single horizontal line
  Baseline - 171{\AA} & 20.3 & 0.66 & 5.2\% & -1.2\% \\
  Baseline - 193{\AA} & 23.6 & 0.68 & 3.4\% & -1.1\% \\
  Baseline - 211{\AA} & 33.0 & 0.84 & 0.9\% & -0.3\% \\
  \hline
  SuNeRF - 171{\AA}  & 34.7 & 0.95 & 0.9\% & -0.3\% \\
  SuNeRF - 193{\AA}  & 43.1 & 0.98 & 0.3\% & 0.05\% \\
  SuNeRF - 211{\AA}  & 43.3 & 0.98 & 0.3\% & 0.02\% \\

\hline                                   %inserts single line
\end{tabular}
\end{table}

\subsection{Complete observation of the solar poles in EUV}
\label{sec:poles}

We evaluate our method using multi-viewpoint EUV observations from two different space missions, SDO and STEREO, with the latter mission consisting of two twin satellites, STEREO-A and STEREO-B.
%the Solar Dynamic Observatory (SDO) equipped with the Atmospheric Imaging Assembly instrument \citep[AIA: ][]{lemen2012aia}, and the twin-satellites STEREO-A and -B each equipped with the EUV Imager \citep[EUVI: ][]{wulser2004euvi}. 
All three spacecrafts orbit in the ecliptic plane, with SDO in geosynchronous orbit around Earth and the STEREO spacecraft leading (STEREO-A) and trailing (STEREO-B) Earth. Even for constellations where we have a full 360$^{\circ}$ coverage of the Sun, we find that three simultaneous images are insufficient for our model training. Therefore, we make use of the solar rotation that allows us to have a full scan of the Sun in $\sim 27$ days, even for a single instrument. For our model training we consider a sequence of 14 days at 1 hour cadence (from 2012-08-24 00:00 to 2012-09-07 00:00 UT; total of 681 observations). We explicitly encode time as a coordinate for the sampling along light rays (Fig. \ref{fig:method}), which allows SuNeRFs to model temporal changes. We randomly sample 32,768 rays within the volume that we then use in our training set. Our model is trained for $\sim 250,000$ iterations (3 epochs) until the model converges (about 2 days on 8 $\times$ A100 GPUs). For our analysis, we primarily focus on the 193 $\text{\AA}$ channel, but we also provide reconstructions of the 171, 211 and 304 $\text{\AA}$ channels. Before training we adjust instrumental differences, including differences in the wavelength bands, using Instrument-To-Instrument translation \citep{jarolim2022instrument}, allowing us to generate homogeneous observation of SDO and STEREO.

\begin{figure*}[!h]
    \centering
    \includegraphics[width=0.65\linewidth]{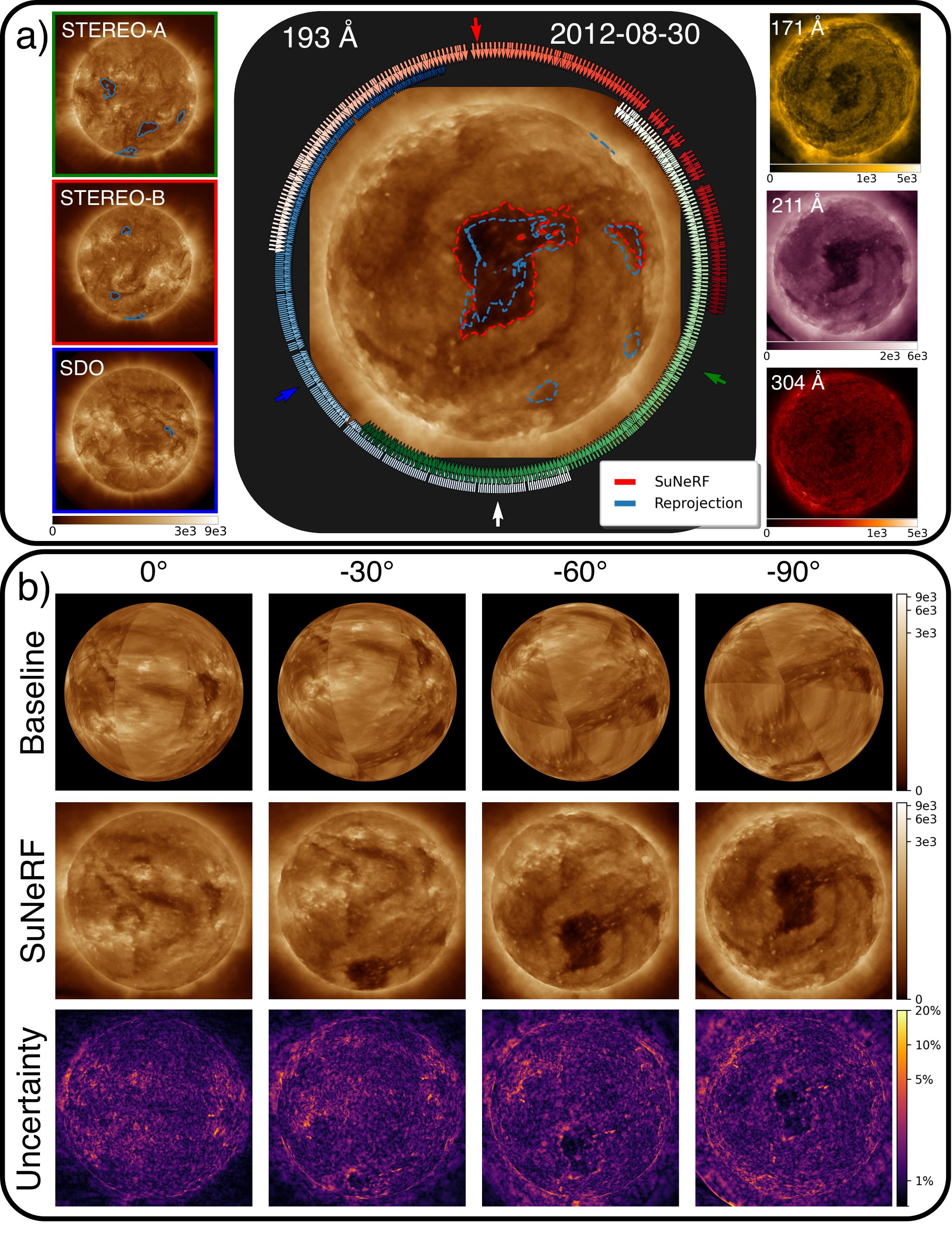}
    \caption{Global reconstruction of the solar EUV corona on 2012-08-30 00:00:00 (UT). (a) Complete image of the solar South pole as reconstructed by our method from STEREO/EUVI and SDO/AIA observations (left). Reconstructions in 171, 211 and 304~$\text{\AA}$ are shown on the right. The image at the center shows the reconstruction in 193~$\text{\AA}$, where the polar coronal hole is best visible. Coronal hole boundaries detected using CHRONNOS are indicated as contour lines: Blue lines refer to spherical reprojections from the ecliptic perspective, and the red lines are obtained from the SuNeRF reconstruction. The arrows indicate the positions of individual observations used for training, and the color coding refers to the temporal distribution of each instrument (green: STEREO-A, red: STEREO-B, blue: SDO). (b) Observations rendered for different latitudes and a fixed longitude (white arrow in (a)). The baseline spherical reprojection (observations stitched into a synchronic map; first row) shows artifacts, provides poor performance for overlapping observations, and is unable to resolve the polar coronal hole. The SuNeRF method (second row) provides consistent observations at all latitudes. The uncertainty maps (third row) indicate regions where higher model errors are to be expected. The EUV colorbars are given in units of DN/s, and the 193 {\AA} images use the same data scaling throughout. An animation highlighting multiple viewpoints can be found in Supplementary Movie 1.} 
    \label{fig:pole}
\end{figure*}

In Fig. \ref{fig:pole}, we show snapshots from a global reconstruction of the solar atmosphere. The arrows in Fig. \ref{fig:pole}(a) indicate the viewpoints used for the reconstruction, with latitudes ranging from $-4^{\circ}$ to $8^{\circ}$. The central image shows the reconstruction of the solar South pole at 2012-08-30  00:00:00 (UT). This directly enables us to further investigate the polar coronal hole, which we segment using a deep learning tool for automatic coronal hole detection \citep[CHRONNOS: ][]{jarolim2021chronnos}. We then compare two approaches in identifying the coronal hole boundaries: (1) We use the individual satellite observations (i.e., full-disk images) to identify coronal holes and then reproject the combined binary maps to the polar viewpoint (blue contours); (2) We use SuNeRF to reconstruct the polar observation and then apply the detection method (red contours). The comparison shows that the strong projection effects only allow for a partial identification of the coronal hole boundary from the ecliptic plane. Coronal holes are also frequently obscured by loop systems in adjacent active regions, which makes detection and consequent space-weather predictions challenging. Our method can account for these projection effects by mapping the 3D geometry (see Supplementary Movie 1). From our comparison to simulated coronal holes, the SuNeRF reconstruction provides a more consistent reference to estimate the coronal hole boundary , specifically in the case of strong reprojection effects, where errors are expected to be in the range of the uncertainty maps (Fig. \ref{fig:pole}(b)). 

In panel (b), we compared viewpoints from different latitudes against the corresponding baseline reprojections. SuNeRF clearly recovers the polar coronal hole by combining the multi-instrument data. 
Artifacts occur only for the off-limb region that is not sampled by the ray tracing (i.e., bottom left at $-90^{\circ}$).

In the bottom row of Fig. \ref{fig:pole}(b), we provide uncertainty estimates derived via an ensemble of five individual runs, performed with random initializations. The uncertainty maps serve as a proxy for the model error, in absence of a ground truth \citep{lakshminarayanan2017simple}. The uncertainty is generally higher for observations than for the model data in Sect. \ref{sec:simulation}. This uncertainty likely arises from the increased complexity of the observations (i.e., the high-frequency features are more difficult to match) compared to the smoother atmosphere in the simulations. Note the higher number of small regions (e.g., bright points) with errors $>10\%$. These could be associated with a misalignment in temporal evolution, where the 1-hour temporal resolution may be too coarse to capture small brightness variations. Overall, large scale features (i.e., active regions, coronal holes, filaments) are consistently reconstructed by our method as can be gauged by the uncertainty maps and the main errors are primarily related to pixel-wise shifts (e.g., coronal hole boundary).

\subsection{Height profiles}
\label{sec:height}

\begin{figure*}[t]
    \centering
    \includegraphics[width=\linewidth]{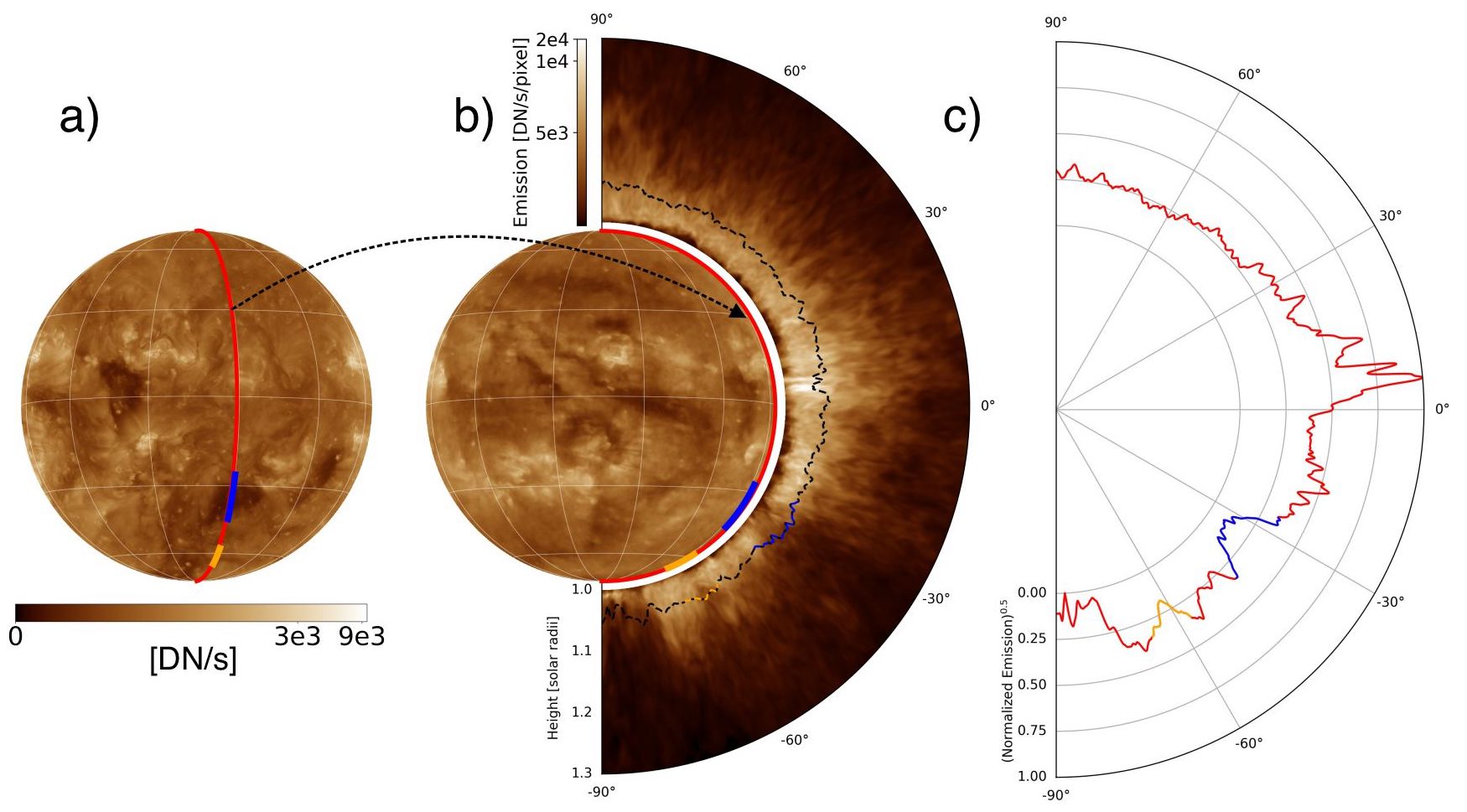}
    \caption{Emission profile of the solar corona as estimated by SuNeRFs. (a) Inter-calibrated STEREO-A observation, where the red line indicates the analyzed slice through the solar atmosphere (fixed longitude). The blue and orange segments indicate the coronal hole and the filament channel, respectively. (b) SuNeRF rendered view, rotated such that the atmospheric slice in (a) coincides with the limb. 
    % We compute a 2D slice of the emission at each height above through the solar atmosphere by calculating the emission at each point. 
    In the solar atmosphere above the said slice, we compute the emission at each height.  
    The 2D atmospheric profile clearly shows the bright active region close to the equator. The coronal holes (blue line and South pole) show a strong decrease in intensity.  The black dashed line indicates the radial averaged height of channel-specific emission, which varies between 15 and 50 Mm ($0.02$ to $0.07\,R_{\odot}$). (c) Square root of the normalized integrated channel-specific emission in the analyzed slice. The emission refers to the radially outgoing total emission, which we use as proxy for the plasma density distribution. The blue segment shows the exact coronal hole profile. The orange segment is associated to the solar filament and appears as sharp drop in the profile (coronal cavity).}
    \label{fig:profile}
\end{figure*}

Using SuNeRFs, we can now determine the height of solar features from the 3D reconstructions. In Fig. \ref{fig:profile}(a), we extract a meridional slice to analyze the emission profile from 1.0 to 1.3 $R_{\odot}$. From North to South, we observe the emission profile of an active region, a coronal hole (blue), a filament channel (orange) and the polar coronal hole. The 2D coronal slice in Fig. \ref{fig:profile}(b) shows regions of lower emission close to the surface. A possible explanation for the modeled emission profile is that our model accounts for the solar chromosphere and transition region, which are not visible in EUV due to temperatures being below the million Kelvin range, to generate EUV emission. In contrast, bright active region loops are rooted in the solar photosphere. The coronal hole can be clearly identified from the decreased emission of the entire column. The solar filament channel, can be identified from the dark cavity which appears as a distinct and sharp drop in intensity. Lower layers show a similar brightness as the quiet-Sun corona, where the reduced emission primarily originates from the dark filament channel. The most distinct emission decrease occurs at the solar South pole, where we observe the polar coronal hole (Sect. \ref{sec:poles}). From the emission we can estimate the relative plasma density by assuming that the emission is proportional to the squared density. This provides a direct estimate of the coronal hole boundaries, which appear as distinct jumps in the integrated emission profiles for the polar and low-latitude coronal holes (Fig. \ref{fig:profile}(c)).

\subsection{Reconstruction of solar eruptive events}
\label{sec:eruption}

To investigate solar transient events, we increase the temporal cadence of observations in our training set and focus on a subframe instead of the full field of view. Here, we apply our method to the solar eruption which started at 2012-08-31 19:30 (UT), and we use observations from STEREO-B and SDO in the 304 $\text{\AA}$ channel. We use the parameters learned from the 304 $\text{\AA}$ model in Sect. \ref{sec:poles} as a starting point, which already provides the general geometry and structure of the considered time period. We focus the training on $1024 \times 1024$ pixel subframes centered at the erupting solar filament (Carrington longitude of 90$^{\circ}$ \& latitude -20$^{\circ}$) at a 1-minute cadence, and train our model for an additional 100,000 iterations ($\sim4$ epochs).

\begin{figure*}[!h]
    \centering
    \includegraphics[width=\textwidth, trim={0 0cm 1.25cm 0cm}, clip]{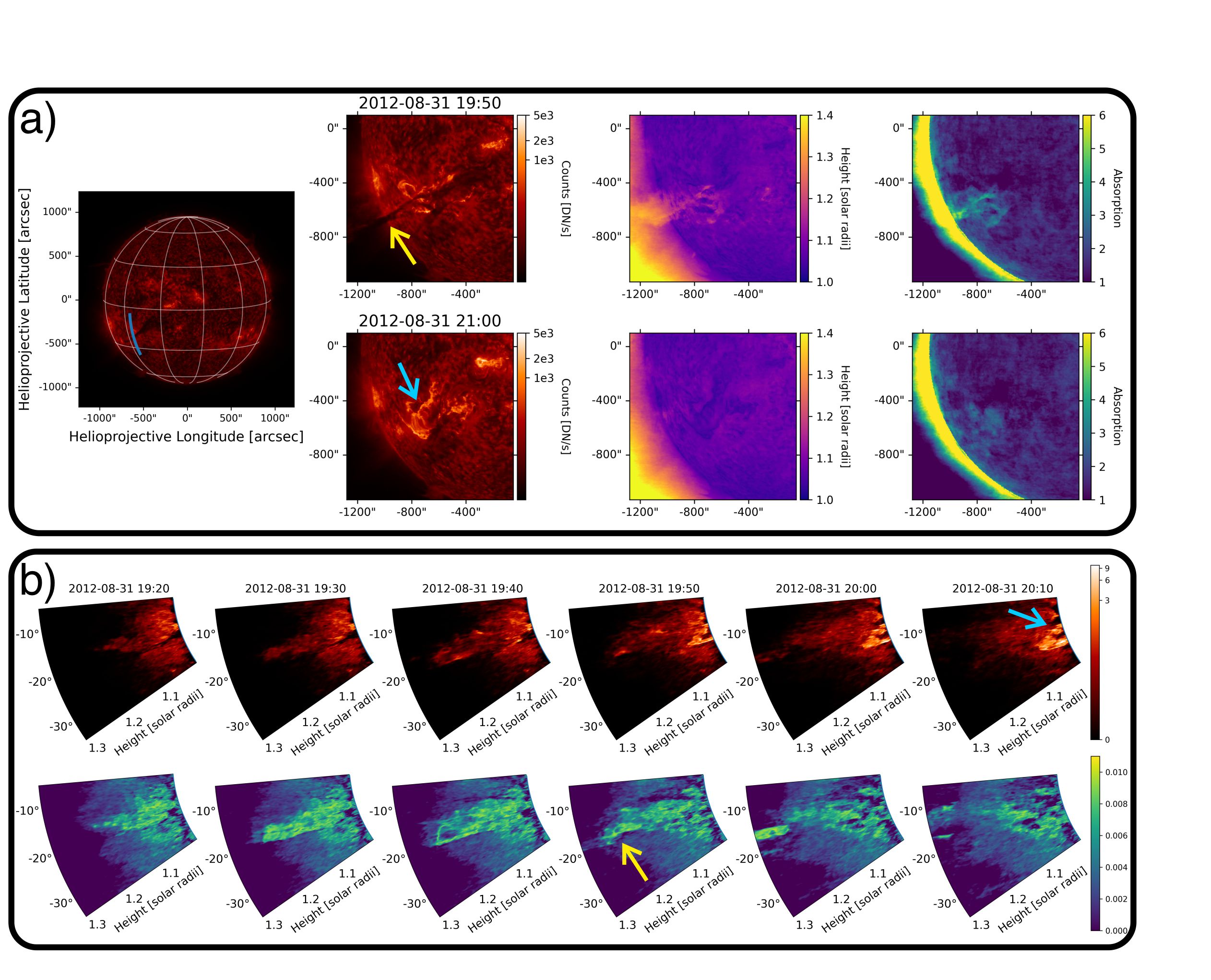}
    \caption{Solar eruption observed on 2012-08-31 in the 304$\text{\AA}$ channel. (a) Full-disk SDO/AIA image before the onset of the eruption, and subframes of our SuNeRF reconstruction. SuNeRF images (left column) are rendered during (top) and after the event (bottom), using the SDO viewpoint. The middle column shows heliocentric height estimates. The right column shows the integrated absorption along the line-of-sight, indicating dense plasma. (b) Temporal evolution of the event as observed for a slice through the center of the solar filament (blue line in panel (a)). We plot the spatial distribution of emission (top) and absorption (bottom) in the atmosphere above the slice. The reconstruction shows a slow rise of the filament, as indicated by regions of strong absorption, and rapid ejection of coronal mass, both visible in emission and absorption. After the ejection, two flare ribbons can be clearly mapped close to the surface (height maps), and post flare loops become visible in the emission slices. See Supplementary Movie 2 for an animated version of this figure.}
    \label{fig:eruption}
\end{figure*}

% In Fig. \ref{fig:eruption}, we show the evaluation of the filament eruption. 
In Fig. \ref{fig:eruption}(a), we give an overview of the filament eruption event, and show SuNeRF-rendered images during and after the eruption (top and bottom row, respectively). We provide height maps that give the distance from the solar center for the observed emission (Sect. \ref{sec:height}). The reconstruction shows that the prominence rises slowly (yellow arrow), followed by a rapid ejection of the filament plasma into interplanetary space (coronal mass ejection). Afterwards, the hot flare ribbons are formed, which are clearly mapped close to the solar surface, in agreement with our physical understanding of eruptions (blue arrow). In addition, we show the total absorption along the line-of-sight, where the filament structure can be clearly identified. The ejection can be related to the depletion of absorbing plasma.

Triangulation is a useful approach for height estimations of dense plasma, but can lead to ambiguous results for an optically thin medium \citep{Aschwanden_2011, Aschwanden2005feature_recognition}. With the full mapping of the solar atmosphere, we can better analyze the underlying 3D structure and study emission and absorption profiles along their radial extent. In Fig. \ref{fig:eruption}(b) we show the time evolution of an extracted slice of the modeled solar atmosphere (blue line in Fig. \ref{fig:eruption}(a)). The top row shows the emission profile. After the eruption ($\sim$ 19:40), post flare loops, and their related flare ribbon footpoints, are formed (blue arrow). The loop height can be directly estimated from the profiles ($\sim 0.7 R_{\odot}$ or 500 Mm above the solar surface). The absorption profiles show the evolution of the filament plasma, and allow to determine the direction and velocity of the erupting structure.

\section{Discussion}

% validity of the method
The analysis in Sect. \ref{sec:results} demonstrates that SuNeRF can use multi-instrument/multi-viewpoint data to create virtual instruments and provide a realistic 3D representation of the Sun. The method is able to reconstruct regions that are only sparsely observed by the individual instruments, like the solar poles. Comparisons to a baseline reprojection demonstrate that our method provides a significant advancement for the synthesis of novel viewpoints. The quantitative evaluation with simulation data shows that our method provides almost identical reconstructions, where differences mostly occur close to the solar limb and in active regions (Sect. \ref{sec:simulation}). Extended regions with less spatial variability (e.g., coronal holes, quiet Sun) are less viewpoint-dependent and thus result in smaller reconstruction errors. 
Uncertainty maps allows us to determine model errors in the absence of a ground-truth reference. For the application to SDO/AIA and STEREO/EUVI data, uncertainties are higher than for simulation data, which we associate with the increased complexity of the corona (e.g., active regions and the solar limb are structurally more complex and detailed than our validation data) and the additional modeled temporal evolution (e.g., the model must now account for the different lifetimes of dynamically-evolving solar features).

% height profiles (advancement)
The SuNeRF reconstructed polar view allows for a more consistent detection of the polar coronal hole boundary than the combined detections from the ecliptic plane (Sect. \ref{sec:poles}), which demonstrates that our method can largely reduce projection effects. With the use of 2D slices through the solar atmosphere, we demonstrate that our method can provide additional insights into solar eruptive events (Sect. \ref{sec:eruption}) and provides a direct estimation of (reprojected) coronal hole boundaries and profiles of solar filaments (Sect. \ref{sec:height}). This provides the basis to directly detect coronal holes from the reconstructed emission profile.

% temporal component
Our method uses video sequences of EUV data for detailed reconstructions with only three simultaneous observing instruments. This is explicitly important for regions that are not covered at a given point in time, but can be approximated from previous and preceding frames. In Sect. \ref{sec:eruption}, we analyzed a solar eruption that is constrained by two viewpoints and reduce the field-of-view while increasing the temporal resolution of the training data. The reconstruction allows to extract additional height information of the event, which can be useful for the estimation of space-weather impacts \citep{thernisien2009cme}. From our model, we can directly map flare related emission (e.g., flare ribbons, loops), and the ejection of dense plasma (i.e., filament eruption). A possible extension that would improve the tracking of coronal mass ejections is the application to coronagraph data that can monitor the propagation through the corona and interplanetary space.

% We note that this reconstruction is enabled through the two viewpoint observations, while we expect that using only a single viewpoint, without additional physical constraints, can not achieve a valid reconstruction.

% only valid results can be expected for the covered region
%We note that our method builds on the reconstruction of a single data series, therefore features that are reconstructed by our method are solely based on the input data, and are not obtained from a generative approach that is aimed to synthesize a realistic appearance.

% limitations + data archive
A limiting factor of our method is the long training time and large computational demand. Training from scratch requires days to reconstruct a single sequence, even on a high-performance computing environment. 
% In App. \ref{sec:solo} we use the trained model from Sect. \ref{sec:poles} as starting point to learn a new sequence. This significantly reduces the training time ($\sim 20$ hours on 4 A100 GPUs for 200,000 iterations). 
For our training, the global structures are learned within the first few 1,000 iterations, but reconstructing details, such as thin coronal loops, requires the full training time. Our method reconstructs sequences of 15 days, which already provides 3D reconstructions in real-time, but individual analysis would require a full simulation run. Extending the temporal range results in a decrease in spatial and temporal resolution, due to the limited capacity of the model and increased training demand. A suitable strategy to reconstruct large data sets would be a continuous simulation, where new frames are added to the end of the sequence while frames at the beginning are truncated and model weights are sequentially extracted. Individual events could then be analyzed by branching and fine-tuning the model with high cadence data of the region of interest. This could provide a higher-level data archive that exploits additional information from the available observations. Note that a larger amount of parameters could potentially capture more details of the 3D volume, but requires more computing power and training time, which is typically the main limitation of our approach.

% similar to tomography, more spatial and temporal resolution, especially important to identify coronal cavities, next step extension to density and temperature profiles which then allows for a direct comparison
Using the multi-channel observations from STEREO/EUVI and SDO/AIA and methods of Differential Emission Measure (DEM), SuNeRF can be extended to give 3D reconstructions of the densities and temperatures in the solar corona. This approach is similar to differential emission measure tomography \citep[DEMT: ][]{vasquez2016corona_3d, vasquez2009multi_tomography, frazin2005rotational_tomography, franzin2009demt}. Here, the SuNeRF approach has the potential to improve the spatial and temporal resolution of the reconstructions, which can be particularly important for the study of faint structures \citep[e.g., coronal prominence cavities; ][]{vasquez2009multi_tomography} and transient events (e.g., EUV waves).

% limitations of coverage
% The analysis presented here uses observations taken with an optimal separation angle between individual spacecrafts, which is ideal for the full coverage of the entire Sun. 
The analysis presented here takes advantage of multi-viewpoint observations captured with an optimal separation angle between individual spacecrafts for a full coverage of the entire Sun.
% In App. \ref{sec:solo}, we apply our method to a narrow separation angle and use only two spacecraft (STEREO-A + SDO). The qualitative comparison shows that our method can produce consistent results for regions that are covered by both instruments. 
The application to a reduced number of viewpoints and the variation in reconstruction quality  could inform future constellation missions (e.g., required coverage). The recent Solar Orbiter mission \citep[SolO: ][]{mueller2020solo} provides additional EUV observing capabilities \citep[Extreme-Ultraviolet Imager (EUI): ][]{rochus2020eui} at non-ecliptic angles that can valuable in further verifying the SuNeRF reconstructions (e.g., by inferring the SolO viewpoint and comparing it to actual SolO observations ) or to extend the training data (e.g., including an additional viewpoint). We note that this requires an additional preprocessing to inter-calibrate the SolO/EUI and SDO/AIA instruments.

% summary + greater impact
Our method achieves a 3D representation of the atmosphere of our closest Star by adapting Neural Radiance Fields. This successful application has implications for other domains in astrophysics and geophysics. NeRF-based algorithms could provide improved 3D reconstructions from planetary, lunar and asteroid surveys, even for complex atmospheric environments (e.g., clouds, dust, gas); all of which are domains likely to soon need such methods as multi-point observations become more commonplace  (e.g., cubesats). A future extension of our method is the use of more physical constraints, with emission and absorption deduced from density and temperature profiles. This provides an additional opportunity to include physics-informed losses \citep{raissi2019pinns} and to connect plasma density with magnetic fields through magneto-hydrodynamic equations \citep{jarolim2022nf2}.

%% IMPORTANT! The old "\acknowledgment" command has be depreciated. It was
%% not robust enough to handle our new dual anonymous review requirements and
%% thus been replaced with the acknowledgment environment. If you try to 
%% compile with \acknowledgment you will get an error print to the screen
%% and in the compiled pdf.
%% 
%% Also note that the akcnowlodgment environment does not support long amounts of text. If you have a lot of people and institutions to acknowledge, do not use this command. Instead, create a new \section{Acknowledgments}.
\begin{acknowledgments}
This work has been enabled by the Frontier Development Lab (FDL\footnote{Frontier Development Lab page: \url{https://frontierdevelopmentlab.org}.}). FDL is a collaboration between SETI Institute and Trillium Technologies Inc., in partnership with NASA, Google Cloud, Intel, NVIDIA and many other public and private partners. 
Any opinions, findings, and conclusions or recommendations expressed in this material are those of the author(s) and do not necessarily reflect the views of the National Aeronautics and Space Administration
The authors would like to thank the FDL organizers, the SETI institute, Trillium, the FDL partners and sponsors, and the reviewers for their constructive comments during the research sprint. 
The authors also want to thank Yarin Gal and Chedy Raissi for valuable discussions. Finally, the authors also thank Google and NVIDIA for providing access to computational resources without which this project would not have been possible. The authors acknowledge the financial support by the University of Graz. A.V. was also supported by NASA grant 80NSSC19K1261 and the STEREO program support to FDL. B.T was also supported by NASA grants 80NSSC21K1796 and NNG04EA00C.
% This research has made use of SunPy v3.0.0 \citep{sunpysoftware2020, sunpycommunity2020}, AstroPy \citep{2013A&A...558A..33A}, PyTorch \citep{pytorch2019_9015}. 
\end{acknowledgments}

%% To help institutions obtain information on the effectiveness of their 
%% telescopes the AAS Journals has created a group of keywords for telescope 
%% facilities.
%
%% Following the acknowledgments section, use the following syntax and the
%% \facility{} or \facilities{} macros to list the keywords of facilities used 
%% in the research for the paper.  Each keyword is check against the master 
%% list during copy editing.  Individual instruments can be provided in 
%% parentheses, after the keyword, but they are not verified.

\vspace{5mm}
\facilities{SDO (AIA), STEREO (EUVI)}

%% Similar to \facility{}, there is the optional \software command to allow 
%% authors a place to specify which programs were used during the creation of 
%% the manuscript. Authors should list each code and include either a
%% citation or url to the code inside ()s when available.

\software{Astropy \citep{2013A&A...558A..33A,2018AJ....156..123A},  
          Sunpy \citep{sunpycommunity2020, sunpysoftware2020},
          PyTorch \citep{pytorch2019_9015}.
          }

%% Appendix material should be preceded with a single \appendix command.
%% There should be a \section command for each appendix. Mark appendix
%% subsections with the same markup you use in the main body of the paper.

%% Each Appendix (indicated with \section) will be lettered A, B, C, etc.
%% The equation counter will reset when it encounters the \appendix
%% command and will number appendix equations (A1), (A2), etc. The
%% Figure and Table counter will not reset.

\appendix

\section{Extended evaluation of simulation data}
\label{sec:extended}

\begin{figure}
    \centering 
    \includegraphics[width=\linewidth]{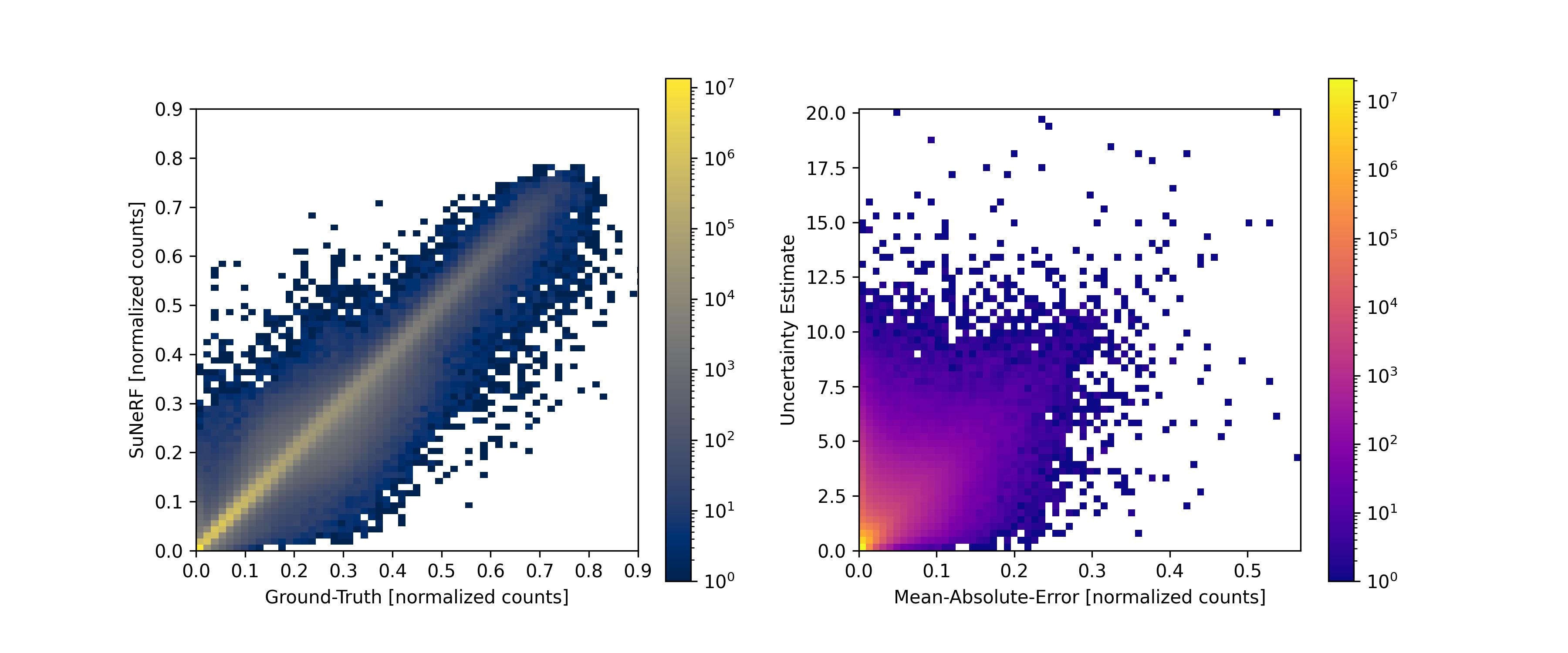}
    \caption{Correlation between SuNeRF reconstruction and ground-truth reference. Left: 2D histogram of simulation data (ground-truth) and our SuNeRF reconstructions. Right: 2D histogram of mean-absolute-error and model uncertainty. The colorbars give the total number of pixels per bin.}
    \label{fig:correlation}
\end{figure}

In Fig. \ref{fig:correlation} we show a comparison between the test set of the simulated data and our SuNeRF reconstructions. We perform a pixel-wise comparison, that shows a clear correlation between the SuNeRF values and the ground-truth reference. The uncertainty estimate and mean-absolute-error have a different scaling, but clearly show a linear correlation, in agreement with the correlation coefficients in Sect. \ref{sec:simulation}.

\section{Synchronic maps}

\begin{figure*}[ht]
    \centering
    \includegraphics[width=0.75\linewidth]{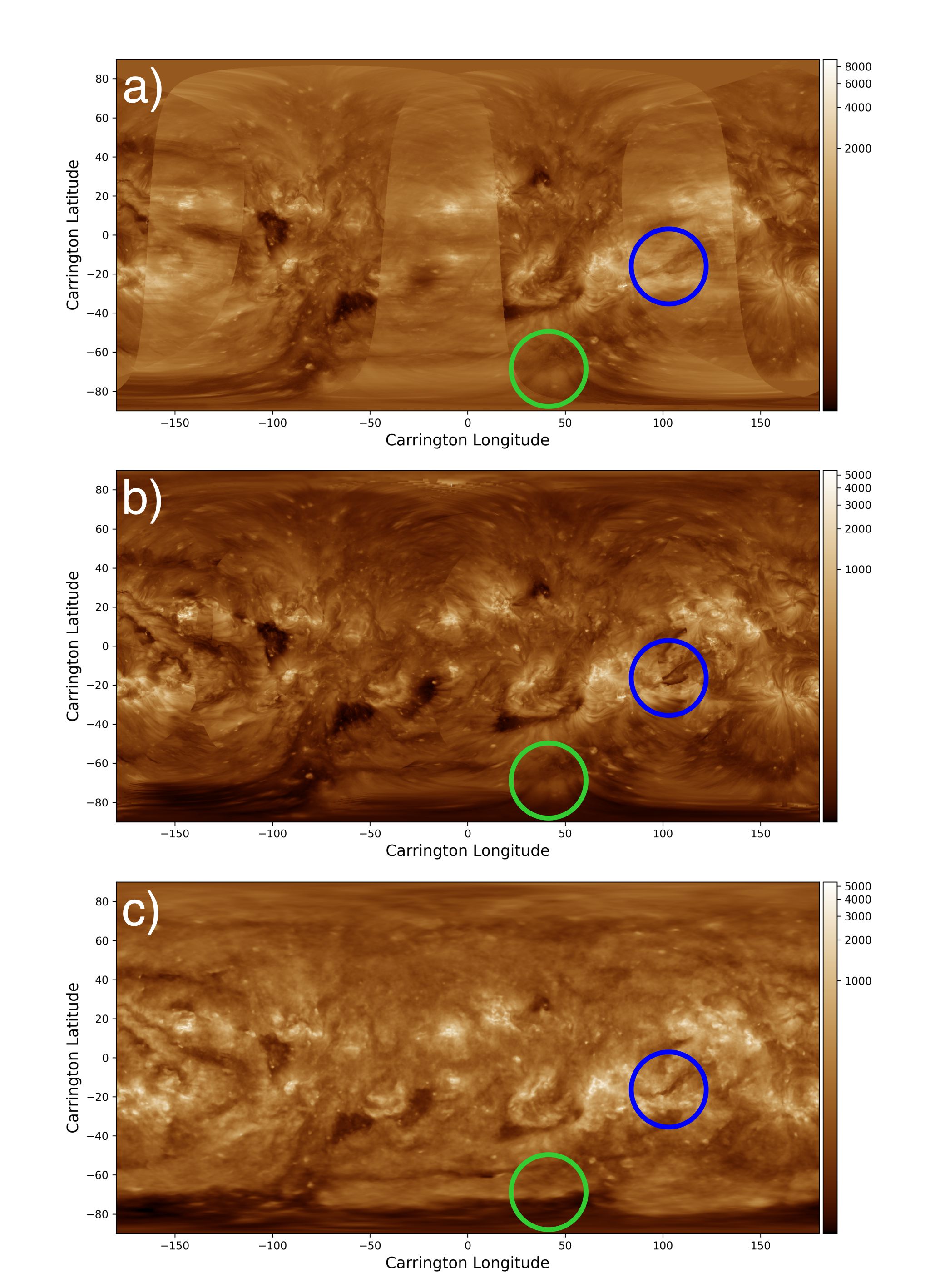}
    \caption{Comparison of synchronic maps on 2012-08-30 00:00 (UT). (a) Conventional synchronic map combining three simultaneous observations from different viewpoints using the reproject function by SunPy. (b) Adjusted synchronic map optimized for coronal hole detections \citep[see ][]{caplan2016}. (c) SuNeRF synchronic map. Reprojected maps can result in poor alignment of features, while our SuNeRF approach consistently combines the three observations (blue circle). Reprojection effects at higher latitudes are corrected by our approach, leading to a clearer identification of the polar coronal hole boundary (green circle). Note that the resolution of our SuNeRF reconstruction is reduced as compared to the synchronic maps, which is related to the limited capacity of the neural representation. The values are given in units of DN/s}.
    \label{fig:synchronic}
\end{figure*}

Using full 3D reconstructions of the Sun, we can render true synchronic maps where each pixel corresponds to an observation normal to the solar surface. In other words, we can obtain ideal reprojected synchronic maps. For this, we sample 512 points from 1.0 to 1.3 $R_{\odot}$ and compute the observed pixel intensity analogous to the ray tracing in Sect. \ref{sec:method}. We sample a full Carrington map ($360^{\circ} \times 180^{\circ}$), where we use a pixel size of $0.025^\circ \times 0.025^\circ$ per pixel. In Fig. \ref{fig:synchronic}, we compare a conventional synchronic map and an adjusted synchronic map \citep{caplan2016} with our SuNeRF map. The conventional map shows projection effects towards the poles, the combination of overlapping regions leads to blurred results, and unobserved regions are interpolated with the mean value (e.g., North pole). The adjusted map improves overlapping regions, but leads to artifacts where features do not spatially align (blue circle). From our SuNeRF map we obtain consistent reconstructions, where solar features at the disk center are similar to the original observations.

\newpage

%% For this sample we use BibTeX plus aasjournals.bst to generate the
%% the bibliography. The sample631.bib file was populated from ADS. To
%% get the citations to show in the compiled file do the following:
%%
%% pdflatex sample631.tex
%% bibtext sample631
%% pdflatex sample631.tex
%% pdflatex sample631.tex

\bibliography{sample631}{}
\bibliographystyle{aasjournal}

%% This command is needed to show the entire author+affiliation list when
%% the collaboration and author truncation commands are used.  It has to
%% go at the end of the manuscript.
%\allauthors

%% Include this line if you are using the \added, \replaced, \deleted
%% commands to see a summary list of all changes at the end of the article.
%\listofchanges

\end{document}